\documentclass[manuscript]{acmart}

\AtBeginDocument{%
  }

\copyrightyear{2026}

\usepackage{hyperref}
\usepackage[
  type={CC},
  modifier={by-sa}, 
  version={4.0},
]{doclicense}

\begin{document}

\title{Repertoires of Contention: Building computational tools and infrastructures for community resistance to bitcoin mining}


 \author{Marina Johnson-Zafiris}
 \authornote{Both authors contributed equally to this research.}
 \email{mlz32@cornell.edu}
 \author{Owen Marshall}
 \authornotemark[1]
 \email{wom6@cornell.edu }
 \affiliation{%
   \institution{Cornell University}
   \city{Ithaca}
   \state{NY}
   \country{USA}
 }

 \author{Jen Liu}
 \email{jl3835@cornell.edu}
 \affiliation{%
   \institution{Cornell University}
   \city{Ithaca}
   \state{NY}
   \country{USA}
 }

 \author{Joseph Ferdinando}
 \email{jgf94@cornell.edu}
 \affiliation{%
   \institution{Cornell University}
   \city{Ithaca}
   \state{NY}
   \country{USA}
 }

 \author{Steven J. Jackson}
 \authornote{Both authors contributed equally to this research.}
 \email{sjj54@cornell.edu}
 \author{Chris Csíkszentmihályi}
 \email{cpc83@cornell.edu}
 \authornotemark[2]
 \affiliation{%
   \institution{Cornell University}
   \city{Ithaca}
   \state{NY}
   \country{USA}
 }




\renewcommand{\shortauthors}{Johnson-Zafiris and M et al.}

\begin{abstract}
  Recent work in human-computer interaction and allied fields has sought to engage community members as co-designers, facilitators, and decision makers to challenge structural power barriers and knowledge gaps. In this paper, we apply a diverse set of computational techniques to engage and address community concerns surrounding the socio-environmental impacts of the Greenidge Bitcoin mining power plant facility in upstate New York. Through collaborative interactions with community groups, addressing problems ranging from regulatory inertia to thermal water pollution, we aim to facilitate not just the production of new and critical environmental knowledge but also new computationally-supported \textit{repertoires of contention}: means of contesting the unequal relations of knowledge and power that structure and  often determine the outcomes of environmental controversies. We argue such engagements can help support more politically engaged forms of HCI, oriented towards efforts to redress and repair disparities in wealth, knowledge, and influence through strategic sociotechnical interventions and alliances.
\end{abstract}


\maketitle
\doclicenseThis

\section{Introduction}
Drawing on ongoing collaborations with community groups opposed to the environmental effects of "proof of work" (PoW) "mining" of "cryptocurrency"\footnote{These widely-used terms refer, rather misleadingly, to a deliberately inefficient and extremely energy-intensive networked computing process through which various speculative digital assets (Bitcoin being the most prominent example) are randomly awarded and transactionally registered on an append-only ledger referred to as a "blockchain." For the purposes of this paper, it is enough to understand that this sort of crypto-”mining" amounts to an environmentally destructive form of gambling.} in upstate New York, this paper examines the tensions and prospects of human-computer interaction as applied to radical organizing and communication in the environmental sphere. Reflecting on the lessons and limits of earlier citizen science efforts in HCI and allied fields, we explore the role of social computation in building new \textit{repertoires of contention}: tactics for the cultivation and deployment of supplemental and adversarial environmental knowledges in worlds marked by weakened regulatory systems, asymmetrical resource distribution, and deep inequities in production of and access to knowledge. In doing so, we argue for HCI’s role in expanding and rebalancing both the cast of actors and the epistemic terrains of environmental controversy, while supporting emergent (re)configurations of collaborative expertise that can build new, if still speculative, relations between science, computation, and community “matters of concern”\cite{latour_why_nodate}. 

The paper that follows combines interview-based ethnography with critical technical practice \cite{agre_toward_2006} in a series of prototypical interventions into public discussions and controversies surrounding environmental impacts associated with the Greenidge Bitcoin-mining facility on Seneca Lake. We open with a review of existing work at the intersection of HCI, scientific collaboration, and citizen science, segueing into examples of community/civic science and data activism research in which political groups and organizations apply computational tools and techniques to entrenched environmental controversies. We then explain the history and background of the Greenidge facility, along with the growing regulatory controversies and community concerns that it has generated. Next, we describe three technical interventions - DEC\_Counts, SatHelper, and GANDER - that we have developed in collaboration with community partners to address concerns around thermal and air pollution  by Greenidge operations. We conclude with a discussion of repertoires of contention – a concept meant to extend HCI’s long standing interest in critical and collaborative technical interventions, especially in the context of environmental knowledge and controversy.

\section{LITERATURE REVIEW}

\subsection{Citizen Science and HCI}

A growing body of research in HCI has sought to support and enhance collaborative work around environmental monitoring, expert accountability, and community engagement in the interest of more just and effective environmental knowledge. An important part of this work has sought to build new collaborations among disciplinary and other credentialed experts, exploring how new computational tools, networks, and infrastructures in fields ranging from ecology and the earth sciences to climate change and species conservation might extend the depth, precision, and (inter)disciplinary and temporal reach of environmental knowledges. This has included work on the the essential role of ‘human infrastructure’ in computationally-driven programs of cyberinfrastructure investment \cite{lee_human_2006}; the crucial work of standardization in establishing commensurability across disparate sites of environmental collection \cite{karasti_infrastructure_2010, millerand_metadata_nodate}; and the deeply affective and value-driven nature of collaborative work in ecology and the earth sciences, even (perhaps especially) in its most ‘technical’ dimensions \cite{jackson_infrastructure_2013,steinhardt_instrumented_2018}. Other work has sought to extend beyond traditional disciplinary experts to draw new classes of “amateur” participants into the making and practice of science. Drawing on celebrated examples from astronomy \cite{raddick_galaxy_2010}, climate warming \cite{eveleigh_i_2013}, molecular genetics \cite{curtis_motivation_2015}, and ornithology \cite{karasti_infrastructure_2010}, and parallel bodies of HCI work in crowdsourcing, gamification, and “human computation” \cite{von_ahn_games_2006}, “citizen science” efforts have attempted to enroll science enthusiasts, amateurs, and non-practicing professionals into the collection and processing of data in large-scale scientific projects.

While these efforts to engage publics in scientific and environmental knowledge production have shown some important successes (see, inter alia, \cite{alvarez_evolution_2020, mckinley_citizen_2017, peters_use_2015, khelifa_wildlife_2016, steinke_accuracy_2017}, they have also shown important limits. Phillips et. al. \cite{phillips_engagement_2019} have emphasized the need to move ‘beyond data collection’ to wider and more meaningful forms of engagement built around personal relevance, learning and place-based connections. Aoki et. al. \cite{aoki_environmental_2017} have written against an overly-simplified reading of motivations and assumed ‘win-wins’ between scientific interests and lay participation, arguing that the interests  informing citizen science may be complex, multidimensional, and periodically contradictory or orthogonal to the stated goals and orientations of projects. Work by Gabrys \cite{gabrys_programming_2014} has shown how citizen sensing efforts may flatten and reduce the category of citizenship itself, breeding new forms of managerialism and environmentality over more robust land- and place-based modes of engagement, resistance, and struggle. Proponents and long-time practitioners have recently argued for abandoning the term itself, noting that many key participants and contributors to leading projects over the years have in fact \textit{not} been citizens in the political sense, nor granted the rights and protections that citizenship affords \cite{ellwood_citizen_2023}. The neutralization and flattening of data may detract from the tactical dimension of these interactions: many citizens become involved with data for a purpose, but already at the level of data collection there are divergent and conflicting purposes, even between citizens in the same community. Moreover, relations to the land that data anchors vary widely (even within movements that may be allied around short-term goals). If data speak to (and from) matters of fact, they are also embedded in what Bruno Latour \cite{latour_why_nodate} has termed ‘matters of concern’: the deeply felt issues – ‘problems’ in John Dewey’s \cite{dewey_public_2012} classic sense – around which interests and commitments gather, fact-making and expertise are deployed, and the terrain of public choice and action is set, including via the regulatory state.

\subsection{Beyond the Citizen}

The community and civic science movement, with links to environmental justice movements, differentiates itself from citizen science by pointing out that traditionally marginalized and low-resource communities are often under-served by environmental science, and that problem-finding from those communities may yield new forms of investigation, knowledge building, and data usage and dissemination \cite{dunbar-hester_hacking_2019}. This perspective complicates pervasive fact-value distinctions central to the social and natural sciences, and places  interests and values, conflict and contention, at the heart of scientific work. Beyond ‘neutral’ facts and data, community and civic science orient towards political and pragmatic questions such as “how do we form groups that actually work?”; “how do we deliberate contested questions of value?”; and “how do we gain access for excluded groups?” \cite{levine_inclusion_2007}. Civic science seeks to operate as an alternative to ‘private science,’ “bring[ing] the infrastructures responsible for sustaining collective human lives… into a space of public contemplation and accountability” \cite{ensemble_electronic_1996-1}. 

Examples of this work may be found in organizations whose projects utilize computational tools and infrastructures no less generative than the ones engaged in citizen science. The Environmental Data Governance Initiative (EDGI), with its loose affiliation of over 100 members from social scientists and librarians to software engineers and planetary scientists, develops analytic tools and infrastructures that monitor government agencies like the EPA, ensures that public needs are centered in their agendas and practices, and assists those agencies and local constituents by supporting and augmenting regulatory infrastructures (including where these are underfunded and under various stages of capture). PublicLab, formed by activists and researchers in response to the 2010 BP Deepwater Horizon oil spill, has developed low-cost and low-entry community accessible tools to document environmental concerns, including kite and balloon-deployed multi-spectral aerial photography, free map and photo integration software, and a spectrographic platform to analyze pollutants. In addition to collaborative development with activists and community members, PublicLab created playbooks and strategies for understanding the impacts polluters have on communities, and leveraging that knowledge to draw public, legal, and regulatory attention.  Littlesis, a node/edge online platform for visualizing networks of power, was not explicitly a “community science” project, but arose from the same interest in citizen research platforms. Populated by thousands of heterogeneous online researchers, Littlesis allows users to map political power and patronage via “Oligraphs” (from oligopoly). Users, including a large number of journalists and activists, monitor and share their visualizations of these networks: linking, for example, state government officials and cryptocurrency lobbyists \cite{zafiris_marina_greenidge_nodate}.

The work of these organizations complements recent scholarship on data activism, where data is used for political action and social change \cite{schrock_civic_2016}. Data activism efforts have sought to critically rework methods of data collection, analysis, visualization, and dissemination to challenge existing inequalities and power structures \cite{tran_situating_2024, milan_alternative_2016} . This work includes counter-data actions which are “an act of data appropriation, whereby individuals intervene to contest the truth of a dataset and construct their own metrics" \cite{meng_grassroots_2018}. In their research with an affordable housing group in Atlanta, Meng and DiSalvo studied how activists collected their own land parcel data to initiate deed transfer requests as a tactic in expanding their community land trust. As another example, southern Californian communities have used data to challenge narratives around homicide as put forth by the Los Angeles Police Department \cite{currie_conundrum_2016}.  Relatedly,  a growing body of scholarship around data centers (see, inter alia, \cite{hogan_data_2015, brodie_cloud_2021, lally_computational_2022}) surfaces tensions between increasing computational activity and detrimental environmental impacts for local residents. This work seeks to refigure these centers as sites of community resistance, often over concerns over water and energy usage. 

\section{CASE STUDY}

\subsection{Practices \& Methods}
In this section we detail our collaboration with communities affected by Bitcoin mining, specifically Seneca Lake-area residents impacted by Greenidge Generation’s gas-fired crypto-mining facility. Our work started in 2021 as the beginning of what became a two-phase research project. The first phase of research focused on understanding and mapping the controversy as it unfolded. [AUTHOR NAME REDACTED] led this work through both in-person and virtual semi-structured and anonymized\footnote{The anonymization of interviewees is both a practical consideration, given previous instances of harassment experienced by activists, and a condition of this project's Institutional Review Board certification.} interviews (n=19) and participant observations with local community members, activists, and experts involved in the issue. Interview questions primarily addressed the interviewee’s relation to and experience with the Greenidge case and/or the Bitcoin mining controversy in general. Interviews were transcribed, coded for specific subtopics (keywords, permits, organizations, etc.) and the resulting material was arranged into a timeline for more comprehensive analysis. This work provided the necessary material to contextualize community concern regarding Greenidge and served as an integral relationship building mechanism with community members and activist groups. Two distinct threads emerged from this ethnographic work which will be expanded upon in the following sections.

The second phase of our research consisted of designing, prototyping and deploying systems to redress issues that emerge from the aforementioned threads of concern. Beginning in March 2022, we made multiple field visits to Dresden, NY (the community neighboring the Greenidge plant). As with the first phase, we regularly organized and attended Zoom-based meetings, discussions via email, call, and text, and in-person discussions with community members to iterate through (re)designs and coordinate deployments. These were not formalized or exclusive design working groups, but rather an ad-hoc collective of individuals that provided critical input when they were available, all with varying levels of participation and collaboration.\footnote{Note: Within our practice, collaboration was not constrained to a particular type of exchange-- some were with us in boats during site visits, others loaned us their piers and waders, some exclusively relayed feedback via online interactions.} We attempted several interventions (often concurrently) with varying levels of complexity, success, and longevity, including collecting and interpreting water samples for toxicant proliferation and conducting participatory GIS kite aerial techniques to map the water surface temperature. In this paper we elaborate on three of these: DEC\_Counts, SatHelper and GANDER.

\subsection{Background: Greenidge, the DEC, and the Seneca Lake Community}
The Greenidge Generating Station began operating as a coal-fired power plant on the western shore of New York’s Seneca Lake in 1938. The electricity it provided enabled a rapid growth in military-industrial infrastructure around Seneca during World War II, including a munitions depot, a chemical plant, and a submarine research facility. By the early 1970s, however, activists and newly-established regulators such as the state Department of Environmental Conservation (DEC) began raising concerns about the plant’s environmental and health impacts, particularly regarding its smoke and fly ash emissions. In subsequent decades, as agritourism overtook heavy industry as the region's main economic driver, Greenidge became one of the state’s least efficient and most heavily polluting plants. Officially decommissioned in 2008, it was later bought by Atlas Holdings LLC, a Connecticut-based private equity firm, which retrofitted Greenidge to run on natural gas and reopened it in 2017 for occasional use as a grid-supplementing “peaker” plant.

When the price of the speculative digital asset Bitcoin spiked that year, however, Greenidge's business model changed from peak grid supplementation to the energy-intensive process of clearing cryptocurrency transactions. By 2020, the plant’s output was primarily used to power thousands of in-house computer rigs dedicated to mining Bitcoin. Greenidge was just one of many such operations which sprouted up across the American rust belt, where decades of deindustrialization had left ample cheap real estate and infrastructure for a growing wave of “crypto-entrepreneurs” \cite{mantius_cuomo_2018}.

Whereas the supplemental energy Greenidge provided to the grid as a peaker plant had been periodic and small, its new Bitcoin rigs’ energy appetite was round-the-clock and virtually insatiable. The result was an eightfold increase in the plant’s C02-equivalent emissions over the course of a year \cite{mantius_greenidge_2021}. Crucial cooling for the server operations was provided by approximately 140 million gallons of cold water drawn every day from the middle of Seneca Lake, which was ultimately returned, now heated, into the lake via the Keuka Outlet trout stream.

As Bitcoin's popularity grew in the wake of the 2008 financial collapse, the DEC was reeling from a wave of debilitating budget cuts. The state-level policy response to the 2008 crisis had been to impose massive reductions in funding for public services while simultaneously seeking a federal bailout of Wall Street financial firms \cite{bill_meyer_aig_2009}. Resulting austerity policies heavily impacted the DEC in particular, which lost roughly half of its budget and a quarter of its staff. This cost-cutting exacerbated the pre-existing epidemic of "zombie permits": permits which have expired but allow continued operation while their renewal requests are either stuck in regulatory backlog or have decisions undergoing administrative appeal \cite{coplan_zombie_2005}. As the DEC budget shrank, the zombie permit problem continued to grow, with industrial monitoring and accountability largely delegated either to volunteer-driven watchdog organizations or private consulting firms commissioned by the polluting entities themselves.

Greenidge’s resurrection and pivot to Bitcoin drew opposition from a wide network of organizations, including regional branches of national environmental advocacy groups like Earthjustice, local not-for-profit corporations like Seneca Lake Guardian, business groups like the Fingerlakes Craft Beverage Environmental Coalition, and independent investigative journalism operations like Water Front Online. With the COVID lockdowns of 2020, these groups met with increasing regularity via Zoom, leading to extensive inter-group collaborations. In July 2021 they formed a state-level alliance called Stop PoW Crypto NY (SPCNY). Similar groups had been forming across the country, and by April 2022 they joined with SPCNY to form the National Coalition Against Crypto-mining (NCAC). Within its first year, NCAC allied with over 18 affiliate organizations and attracted participants from over a dozen states. It was in this context that SPCNY’s member organizations began organizing specifically against the renewal of Greenidge’s Title V air permit.

\subsection{In the Air}

\begin{figure}[h]
  \centering
  \includegraphics[width=\linewidth]{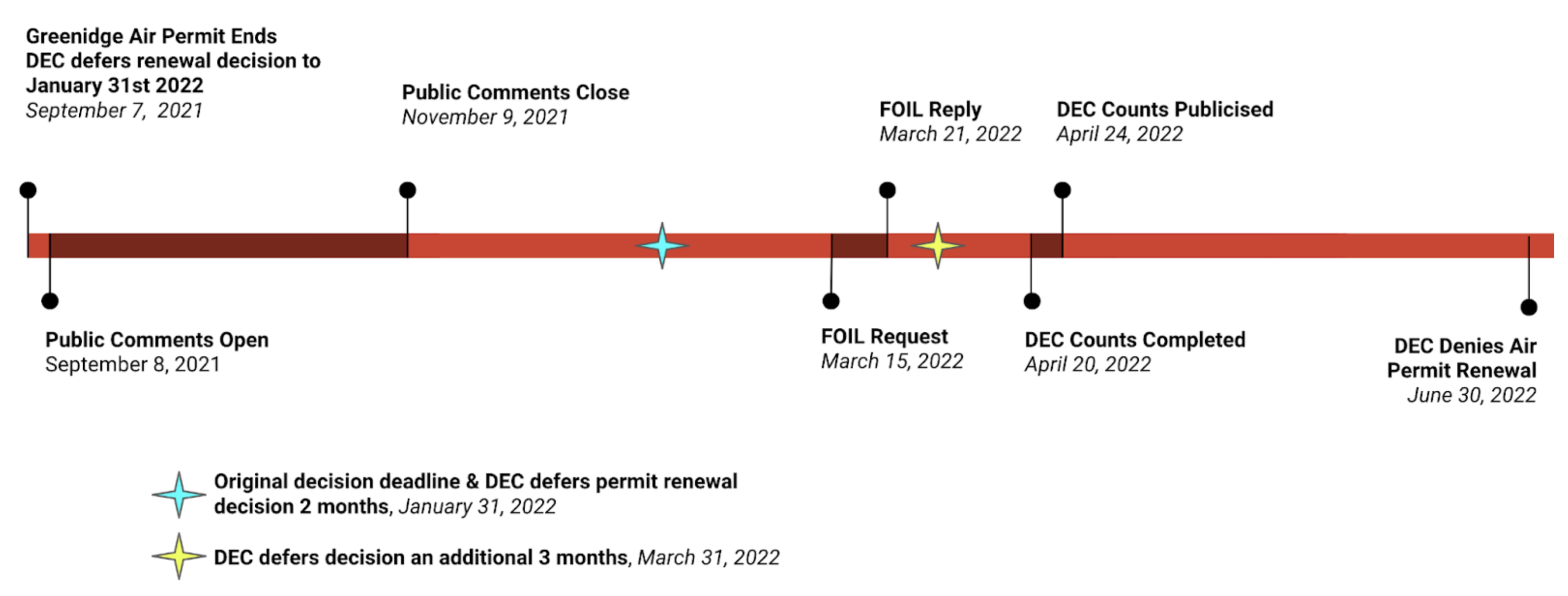}
  \caption{Timeline of SEQRA Title V air permit public comment interaction.}
  \Description{A horizontal timeline of SEQRA Title V air permit public comment interaction.}
\end{figure}

On September 7, 2021, in accordance with the New York State Environmental Quality Review Act, the DEC opened public commentary regarding Greenidge’s Title V air permit renewal. Multiple activists and community members organized public comment campaigns specifically demanding that Greenidge’s application for the permit be denied on the grounds of inability to satisfy Section 7(2) of New York's ambitious greenhouse gas reduction legislation, the Climate Leadership and Community Protection Act (CLCPA). Campaign organizers also encouraged locals to express their own experiences and accounts of environmental deterioration in the aftermath of Greenidge's pivot to Bitcoin mining. As a result, over 102 oral statements were made at the two legislative public comment hearings and over 3,883 written comments were received during the public comment period.

Our first round of interviews coincided with the DEC’s repeated deferrals of their decision regarding Greenidge’s air permit renewal request. (In keeping with the “zombie permit” policy, the DEC had allowed the plant to continue operating on a permit which had expired in September of the previous year, with a renewal decision further postponed so the DEC could “complete its ongoing review” of the roughly 4,000 responses it had received \cite{mantius_dec_2022}.) Interviewees were uniformly skeptical of this justification. As one community organizer put it, the DEC’s official rationale was “Bullshit. I mean, you know, they could finish those comments in a night if they had the will to do so. So we don't feel that that's a very genuine answer. There's politics going on.” There was also skepticism around the officially-promulgated notion that the decision would ultimately be made on scientific grounds by the DEC commissioner, Basil Seggos. In the words of a prominent engineering scientist involved in the issue, “let’s be realistic, the decision is with the Governor… the DEC is supposed to make decisions based solely on the science, okay, but in the real world that’s not how it works… [Governor Kathy Hochul] is being lobbied heavily and she’s being pressured by politicians in her own party.”

\subsubsection{DEC\_Counts}
In response to the DEC’s repeated deferrals (Fig. 1), Seneca Lake Guardian organizers we had connected with through our fieldwork asked our team to analyze and obtain general descriptive statistics of the public comments received (which we had secured earlier, on a USB drive delivered by mail, through a Freedom of Information Law (FOIL) request submitted to the DEC in March 2022).\footnote{The comments, delivered by mail in a USB drive, were presented variously in .pdf, .doc, and in some instances Outlook .msg formats, which complicated streamlined descriptive and quantitative analysis. The preliminary steps of DEC\_Counts thus consisted of data cleaning methods which compiled material into accessible and parsable .txt and .csv formats for analysis.}

\begin{figure}[h]
  \centering
  \includegraphics[width=\linewidth]{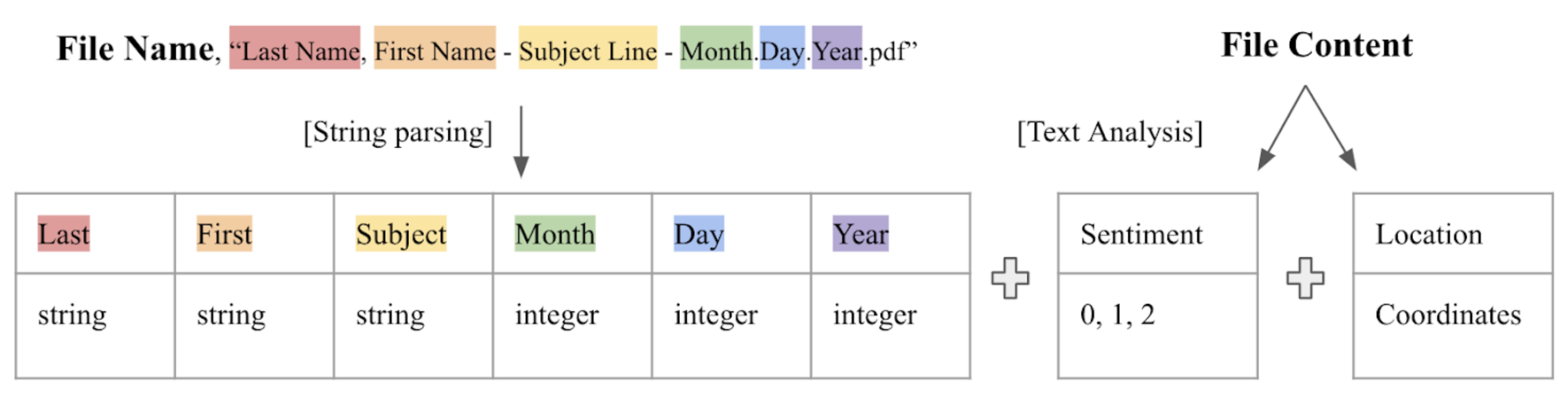}
  \caption{File parsing structure for preliminary DEC\_Counts sentiment analysis.}
  \Description{File parsing structure for preliminary DEC\_Counts sentiment analysis.}
\end{figure}

For the preliminary analysis, we took a quantitative and qualitative mixed methods approach, employing techniques such as automated text analysis through subject string parsing, text annotation and sentiment ground-truthing. Most of the files provided by the DEC followed a “Last Name, First Name-Subject Line- Month.Day.Year.pdf” naming structure as seen in (Fig. 2). A baseline result for “anti-renewal” sentiment was easily established through subject line patterning. The majority of public comments were submitted through community based campaigns with titles like “Save NY from Climate-Killing Cryptocurrency” and “Deny Greenidge's Title V Air Permit.” However, other subject lines were more ambiguous, such as “Greenidge Generation Power Plant” and “Greenidge Comments,” and necessitated qualitative comment content analysis (i.e., close reading). Through our analysis we found that, of the roughly 4,000 PDF and Word files provided, over 98 percent expressed clear opposition to Greenidge's air permit renewal, with less than 1 percent expressing support (Fig. 3). This method was corroborated manually by sorting comments into folders and spot-checking for outliers.

We then georeferenced the public comments with their corresponding sentiment. For this purpose, DEC\_Counts utilizes Named Entity Recognition algorithms to extract the address of specific public comments that have opted-in to disclosing their location. We then abstracted this data into latitude and longitude coordinates and visualized them via Folium data map, in which viewers could zoom-in, zoom-out, hover and click on particular sentiment dots (red representing opposition, green representing support for the permit renewal). This process established the relation between  individual comments and specific places to address common concerns of public comment processes  hijacked or taken over by outside groups. The resulting map exposed the overwhelming local and statewide distribution of anti-renewal sentiment in contrast to the small number of pro-renewal files (Fig. 4).

Once the results of the analysis were complete we presented this material to Seneca Lake community activists, who then publicized the findings through their network of contacts. A press conference was organized with [authors' names redacted] as main presenters; examples of the resulting press are featured below [redacted for review]. 

\begin{figure}[h]
  \centering
  \includegraphics[width=\linewidth]{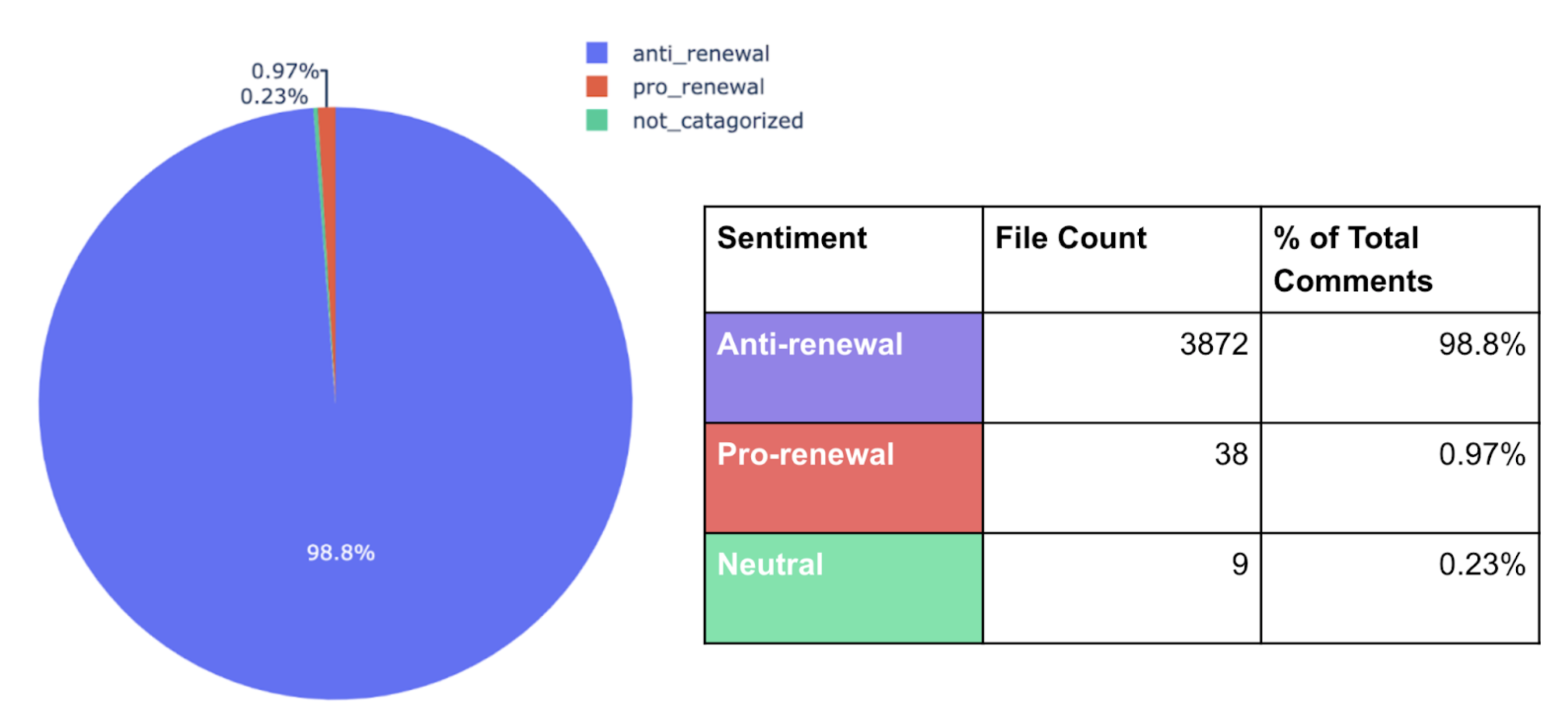}
  \caption{Results from sentiment review of public comments}
  \Description{Pie chart and table with results from sentiment review of public comments}
\end{figure}

Often, the DEC publishes a “Responsiveness Summary” to the public comments themselves, extracting the main points of the comments if there is a large quantity, and writing “Department Responses” which often point back to procedural requirements the DEC adheres to. In the case of Greenidge’s Air Permit Renewal, there was no Responsiveness Summary. But even if there had been, valuable contextualization and metadata of the kind provided in our analysis would have been lost. What was the overall weight and balance of public sentiment? Where were the comments coming from (within or beyond the region? authentic expressions of local sentiment or reflecting the influence of outside actors and interests)? What were the distinct stories embedded in both campaign and non-campaign based messages? Our analysis produced material that the DEC had not offered before and at a pace that allowed activists to demand department response.

\begin{figure}[h]
  \centering
  \includegraphics[width=\linewidth]{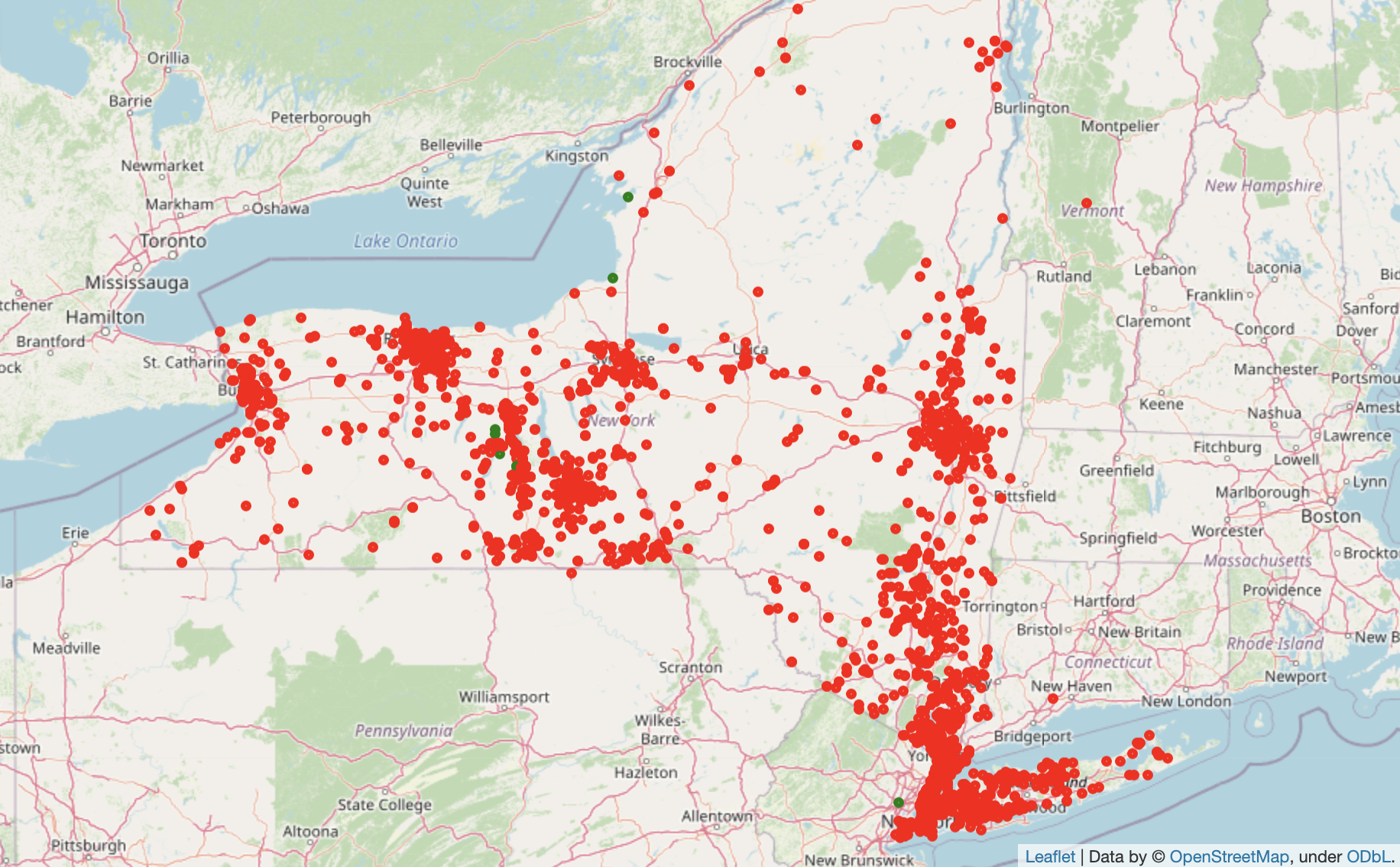}
  \caption{ Sentiment Geomapping, red dots pertain to comments that expressed anti-renewal sentiment and green pertain to comments containing pro-renewal sentiment (neutral comments were excluded)}
\end{figure}

This intervention highlighted the valuable nature of computational sentiment analysis as a mode of 1) challenging the timelines of government information dissemination and aggregating the value of community and activist commentary; and 2) reinforcing community activists' ability to conduct "objective" analysis of a large corpus of "subjective" comments through quantitative methods and tactical machine learning. Our present expansion of DEC\_Counts has manifested in an analytical tool that allows users (primarily environmental activists and community members) to:
\begin{itemize}
    \item Upload public comment datasets
    \item Automate the data cleaning process
    \item Produce descriptive statistics of the general sentiment
    \item Employ unsupervised machine learning tactics, including vectorizer and K-means algorithms, to cluster campaign commentary and extract unique public stories and opinions
    \item Generate data visualizations that convey public opinion more effectively
\end{itemize}
DEC\_Counts also continues to develop, as we explore additional design features that allow users to input a diverse range of public comment datasets. We currently have two on-going applications of DEC\_Counts external to this case study, elaborated further in the discussion section below. 

Influenced in part by these public comments, the DEC did ultimately deny Greenidge’s air permit renewal request, approximately six months after the content of the public comments became widely known–a delay that some observers attributed to Governor Hochul’s interest in waiting until after she had been successfully re-elected \cite{mantius_hochul_2022}. This delay did not kill the permit, however, but only “zombified” it (i.e. turned it into an expired yet still quasi-functional  “zombie permit” as discussed above). As of this writing, Greenidge is suing the DEC to prevent it from enforcing the permit denial. A state supreme court hearing on the issue is scheduled for early November, 2024. 

Public sentiment and related press coverage surrounding the Greenidge Air Permit case also  contributed to the successful passing by the state legislature of a two year moratorium on using fossil fuel power plants as crypto mines. This was an important activist gain, even though Greenidge was grandfathered in and thus largely unaffected by the new law. The moratorium is set to expire on November 22nd, 2024. The moratorium act also required the DEC to produce a statement on the environmental impacts (EIS) of crypto mining. A draft outline of this statement was issued for public comment in early 2024 and finalized in April. A draft of the actual EIS will presumably be issued before the end of November. A final version will be issued followed by a 120 day public comment period. We plan to apply the DEC\_Counts methodology to this new set of comments once they become available.

\subsection{In the Water}
 As controversies surrounding the Air Permit unfolded, Seneca lake community activists also began asking questions around Greenidge’s hot water discharge: both in relation to its regulatory upholding under the State Pollution Discharge and Elimination System (SPDES) and its general impact on the lake ecology. Interviews revealed a similar degree of incredulity regarding the DEC’s handling of the plant’s SPDES water permits, which set limits on temperature and other forms of pollution in the water Greenidge discharges into the lake.

The SPDES Program, launched in the 1970s, is intended to “eliminate the pollution of New York waters and to maintain the highest quality of water possible” \cite{noauthor_state_nodate}. Through a well-funded lobbying effort \cite{mantius_cuomo_2018}, Greenidge had secured an unusually permissive SPDES permit, allowing discharge temperatures of up to 108 degrees Fahrenheit into a trout stream feeding the lake, well above the normal legal limit of 70 degrees. At the launch of our project, Greenidge's SPDES permit was set to expire on September 30, 2022; while Greenidge filed its renewal application in January 2022, at the time of writing in September 2024, the renewal remains unapproved (with the existing permit still operating in ‘zombie’ mode). One of the few burdens the now expired permit placed on Greenidge was the requirement to conduct a thermal study of its water output into the Keuka Outlet. 

One potential impact of thermal water pollution is the growth of harmful algal blooms (HABs). Seneca Lake had long been assumed to be immune to HAB problems because of its status as a deep and cold lake with low nutrient content. This assumption did not hold when a sharp spike in confirmed toxic blooms broke out in 2017 - the same year that Greenidge resumed operations. Concerned that the plant’s hot water emissions might be lengthening the HABs growing season, and frustrated by the DEC’s multi-year deferral of the required thermal study, local community members took it upon themselves to measure the temperature from the outlet and present their findings to the town planning board. These temperature charts indicated inlet temperatures exceeding 100 degrees Fahrenheit: well under Greenidge's 108 degree limit but still far in excess of normal lake temperatures and typical regulatory standards. 

Because of practical and legal issues involved with placing the sensors and keeping them from being tampered with, activists were reluctant to provide details such as the exact placement of the sensors or the names and qualifications of the people involved. This lack of transparency regarding the data’s provenance made it easy for officials to ignore these findings. At the Yates County planning board meeting of September 21st, 2020, where the citizen-gathered temperature data was first made public, community members asked that the proposed expansion of the Greenidge operation be delayed until the two year overdue study of thermal pollution and its potential impacts could be completed. Planning Board Chair David Granzin dismissed the issue, arguing that the permit question was outside the board's jurisdiction because “DEC controls Greenidge permits.” Greenidge did not actually complete the Thermal Discharge Study until May 2022, however, and it was not made available to the public until 2023. The study itself was conducted by a hired third party contractor, ASA Analysis \& Communication and Najarian Associates \cite{mantius_greenidges_2022} .

Given the obscured nature of the Thermal Discharge Study, we collaborated with community members to create alternatives to inspect the thermal state of the lake and outlet discharge zone. They expressed a desire to both replicate Greenidge’s Thermal Discharge Study as well as generate their own baseline understanding of Greenidge’s thermal plume. They also sought to improve on two significant flaws in the earlier community monitoring effort: reliability and scope. On reliability, although community members placed thermal sensors themselves, they were not always able to find or retrieve them, and expressed concern that Greenidge employees were tampering with them. (At the same time, Greenidge expressed concern in reports and presentations that its own sensors were being tampered with, and introduced security cameras to cover some of their sensor locations \cite{asa_analysis__communication_inc_greenidge_2022}). On scope, community activists noted that placing single sensors was both costly and lacked resolution. While there were ways of extrapolating point data, the community members preferred a picture to a pixel. 

Most previously developed citizen-oriented water sensing platforms (e.g., Mayfly and AguaClara) and commercial water thermometers yield point data; the character of the thermal discharge plumes on Lake Seneca – wind-driven, quickly moving, and convective – resists such sampling techniques. This was supported by the DEC itself, when it emailed one scientist who had been recruited by the community to observe that, “given the variable circulation around the Dresden shelf area which affects the direction and size of the river plume into the lake, having just one fixed point of monitoring would make it difficult to assess what is going on.” Additionally, activists were only able to easily place sensors on the shoreline, docks, or the banks of the heated outlet, as floating object permits and the engineering necessary to place permitted buoys were impracticable.\footnote{Personal communication, April 14, 2022.} 

Community-based sensing platforms are sometimes dismissed because they generally fall short in litigation efforts or as evidence for legislation: expert witnesses challenge their data through questions of sensor calibration, chain of custody, and bias concerns. The alternative, hiring third-party bonded science and engineering firms, are often unaffordable. But community concerns follow many stages. In our case, the community was suspicious that the billions of gallons of heated water being discharged into their lake per month were having an effect on its temperature. Sensing platforms during this "hunch" period, while not generating legally admissible data, help communities to understand the challenges they are up against, and may indicate where to direct future attention. Activists, who often initially find themselves excluded from the technical dialog between industry and regulators, determine that they have to "learn the lingo" and bring a technical argument themselves. They recognize the political agency of what Shapin called a "modest witness," a technical apparatus that purports to speak for nature itself, without human opinion, politics, or interests. While we know that such witnesses are never actually neutral, are often manifestations of authority rather than a force in themselves, nor do they resolve scientific controversies on their own, their continued power in disputes remains unassailable.

Based on this community input, our team began experimenting with two complementary prototypes to better trace the heat waste problem: 1) utilizing and supplementing the open access benefits of satellite imagery that yields public, but difficult to access and interpret, data; and 2) deploying automated/mobile/robotic approaches to evaluate sub-surface water temperature in paths (matrices). These proofs of concept, not yet deployed in full, are described in turn below as examples of future possible directions. 

\subsubsection{SatHelper: Making Satellite Imaging More Accessible}
A variety of agencies worldwide collect remote sensing data and release it free for public use. Satellite observations like those by Landsat have low spatial and temporal resolution and take only surface measurements, but represent a unique opportunity to monitor locations without deploying hardware or engaging in extensive and costly fieldwork. Additionally, they can provide a record of past conditions in data archives for the purposes of change detection and measuring longitudinal impact. While it has become commonplace to access satellite and aerial imagery in commercial mapping services, these basemaps are cloud-mosaics that do not reveal date-time metadata and are not always frequently updated. Satellite imagery taken by NASA, The US Geological Survey (USGS), and European Space Agency (to name a few) are updated on daily and weekly tempos, depending on resolution and coverage. This data is, in principle, freely available to the public, though  often stored in databases and formats that are obscure or difficult to work with. Most imagery retrieval platforms require logins, a fair amount of subject matter expertise, and sometimes coding experience to use their data effectively. Images tend to be organized by agency/satellite, further confounded by a single satellite platform or sensor usually serving a variety of functions and creating a number of derived products.

\begin{figure}[h]
  \centering
  \includegraphics[width=\linewidth]{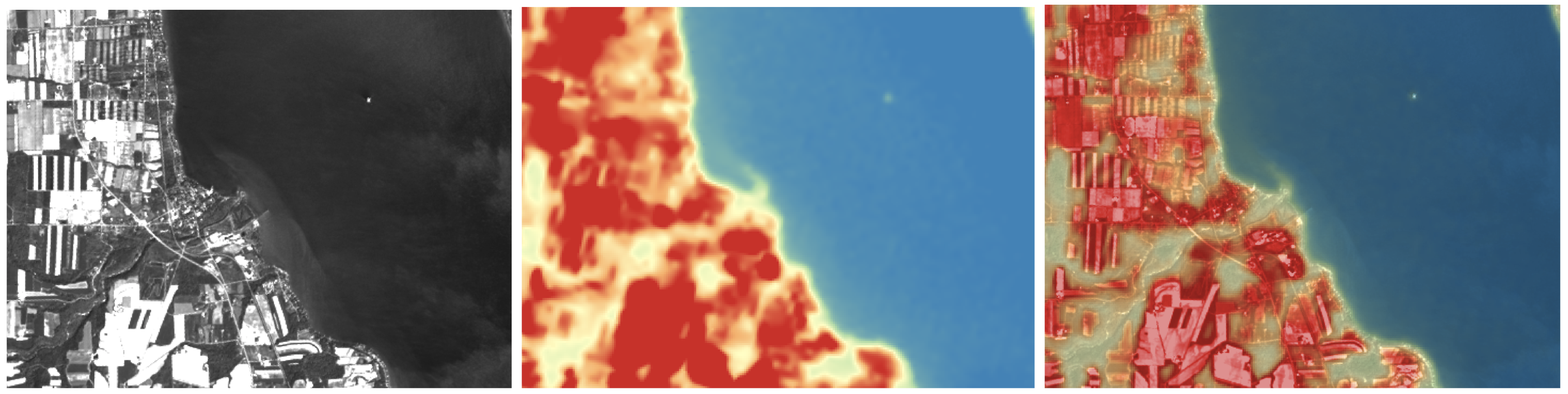}
  \caption{Left: the high resolution panchromatic band shows the context of Seneca Lake; center: long-wave infrared image of the same location at the same time; right: image bands composited to show the size and relative zone of high temperature on the water surface. Note thermal plume from Greenidge via Keuka Outlet in the center of each image.}
\end{figure}

We attempted to meet community interest in Seneca Lake temperature data measurements with streamlined access to otherwise illegible publicly-funded satellite imagery. Here the focus is on surface temperature measurements taken by the Landsat 8 and 9 satellites, though temperature measurements from other sensors (ECOSTRESS, MODIS) might further increase temporal resolution, and other higher resolution optical images might be applied to a variety of conservation and industrial monitoring use cases. These thermal images require the area to be cloud-free, and more reliably show thermal water surface plumes in nighttime images, which luckily are increasingly common operations for the Landsat satellites.

\begin{figure}[h]
  \centering
  \includegraphics[width=\linewidth]{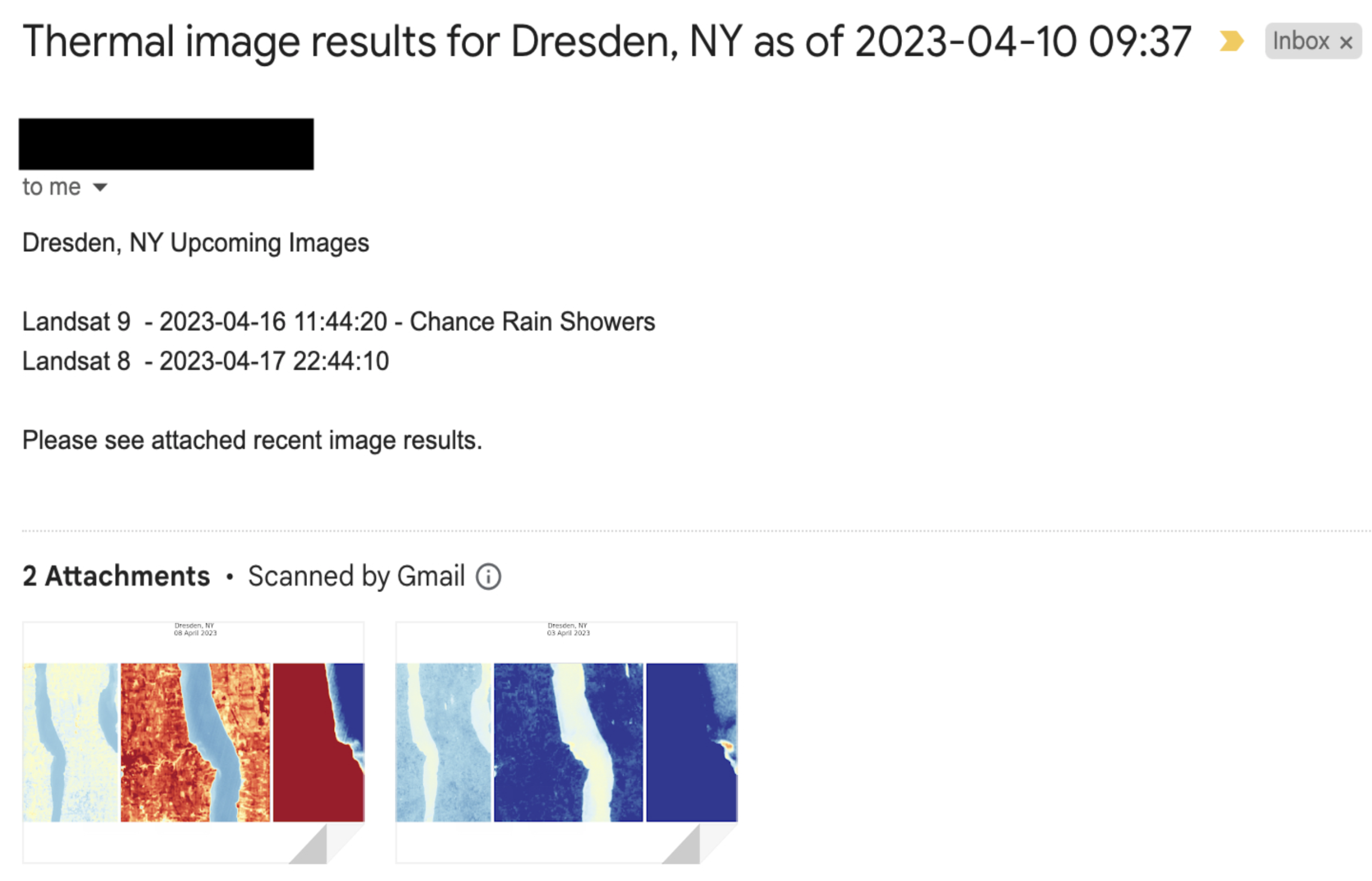}
  \caption{Example email from the SatHelper service.}
\end{figure}

SatHelper provides both information on upcoming overflights, and cropped and color-mapped recent images of a site in question, retrieved from public sources. This information is provided to the subscriber’s email address (eliminating logins), and sent as .jpg or .png attachments rather than large GeoTIFF files that can require geospatial software to properly view. From a latitude/longitude input, SatHelper uses the National Weather Service API to identify the correct NWS station, and retrieves weather forecasts to provide the user with an expectation of cloud cover. (As Landsat images are optical (visible and infrared) their ability to see a subject on the ground is limited by clouds.).  satellite tracking API (REST API v1) with a “radio pass” function that forecasts when a satellite will be visible from a specific ground location and angle is then merged with the relevant weather forecast conditions,  and a Landsat Machine to Machine API to retrieve available images is applied. SatHelper then crops and stretches these images and applies a color scheme at various zoom levels to show both context and detail, and shared as image attachments viewable without any specialized GIS knowledge or software.

\subsubsection{GANDER: Geolocated Aquatic Navigation, Detection, and Evaluation Robot}

\begin{figure}[h]
  \centering
  \includegraphics[width=\linewidth]{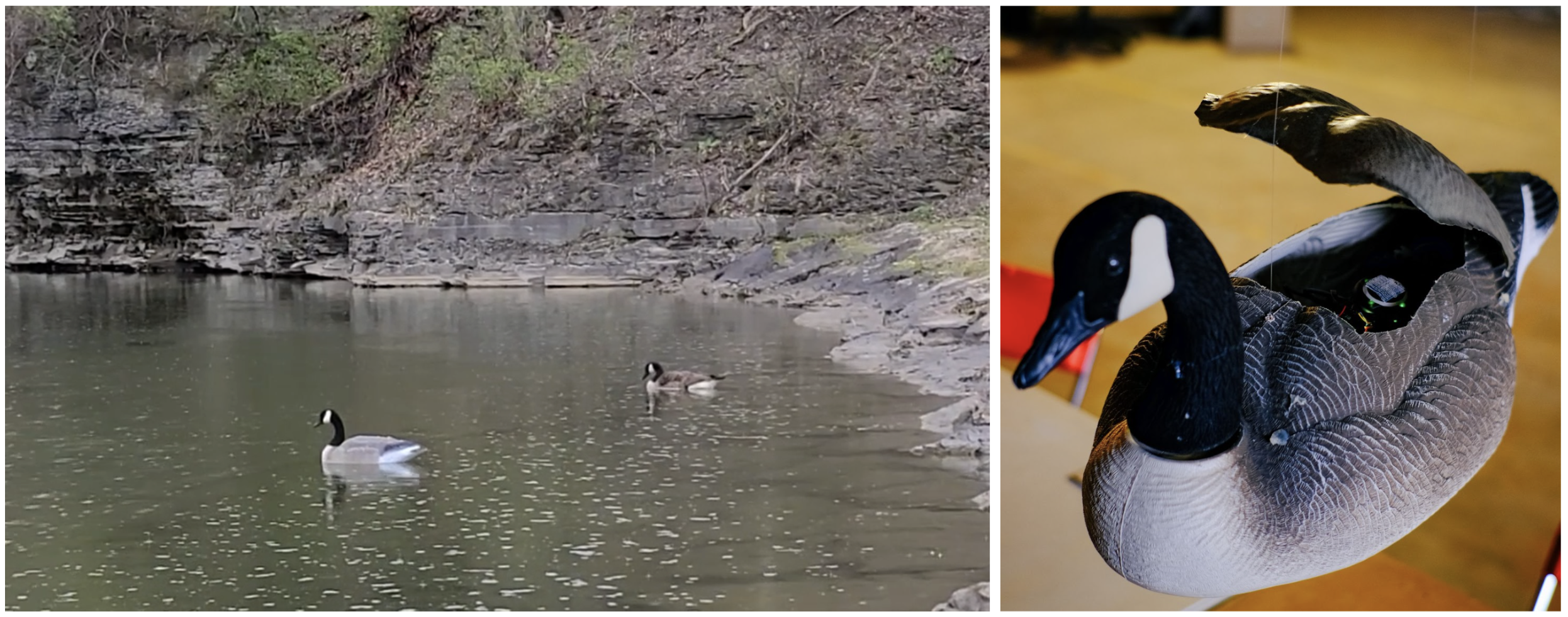}
  \caption{Left image illustrates GANDER test run, "real"  branta canadensis is the one closest to shore. Right image is a close-up of GANDER prototype.}
\end{figure}

The second element of our thermal lake study was to capture surface temperature of the thermal waste plume using a single temperature sensor on a mobile platform. A mobile platform is technically far more complicated than a point temperature sensor, but it bypasses the regulatory and marine engineering hurdles of mounting permanent buoys (requiring permits), and can yield data across a geographic vector or matrix (though not simultaneously to all points in the matrix). At the same time, any easily recognizable platform would be visible to employees of Greenidge or random lake traffic, and thus susceptible to the tampering that both community members and Greenidge claimed was affecting their point temperature sensors. Without coming down on either side, we wished to avoid the possibility of tampering, and strategized clandestine ways of sensing. We drew inspiration from the then popular “Birds aren’t real” meme, and chose to design the mobile sensing platform to appear as a \textbf{branta canadensis}, or Canada Goose, a species endemic to the Finger Lakes region. We accomplished this by designing the platform to fit into the body cavity of a “Big Foot Canada Goose Floater” hunting decoy. The platform, which we nicknamed GANDER, has proven indistinguishable from a live goose to active and casual observers, and confuses (and occasionally alarms) even real geese.

The autonomous surface vehicle uses a popular drone autopilot hardware platform – Pixhawk – using Arduboat firmware. A custom underwater electric prop and rudder mechanism is servo-steered with a Proportional Integral Derivative (PID) control loop, and can autonomously navigate to preset GPS waypoints, or be remotely piloted via a phone or laptop through the 4G network. An onboard RPi4 single board computer handles communications and is interfaced to a Measurement Specialties TSYS01 temperature sensor with .1°C resolution. Measurements are injected into the Pixhawk’s MAVLINK data stream using the Python Mavutil library and MavProxy.  

\begin{figure}[h]
  \centering
  \includegraphics[width=\linewidth]{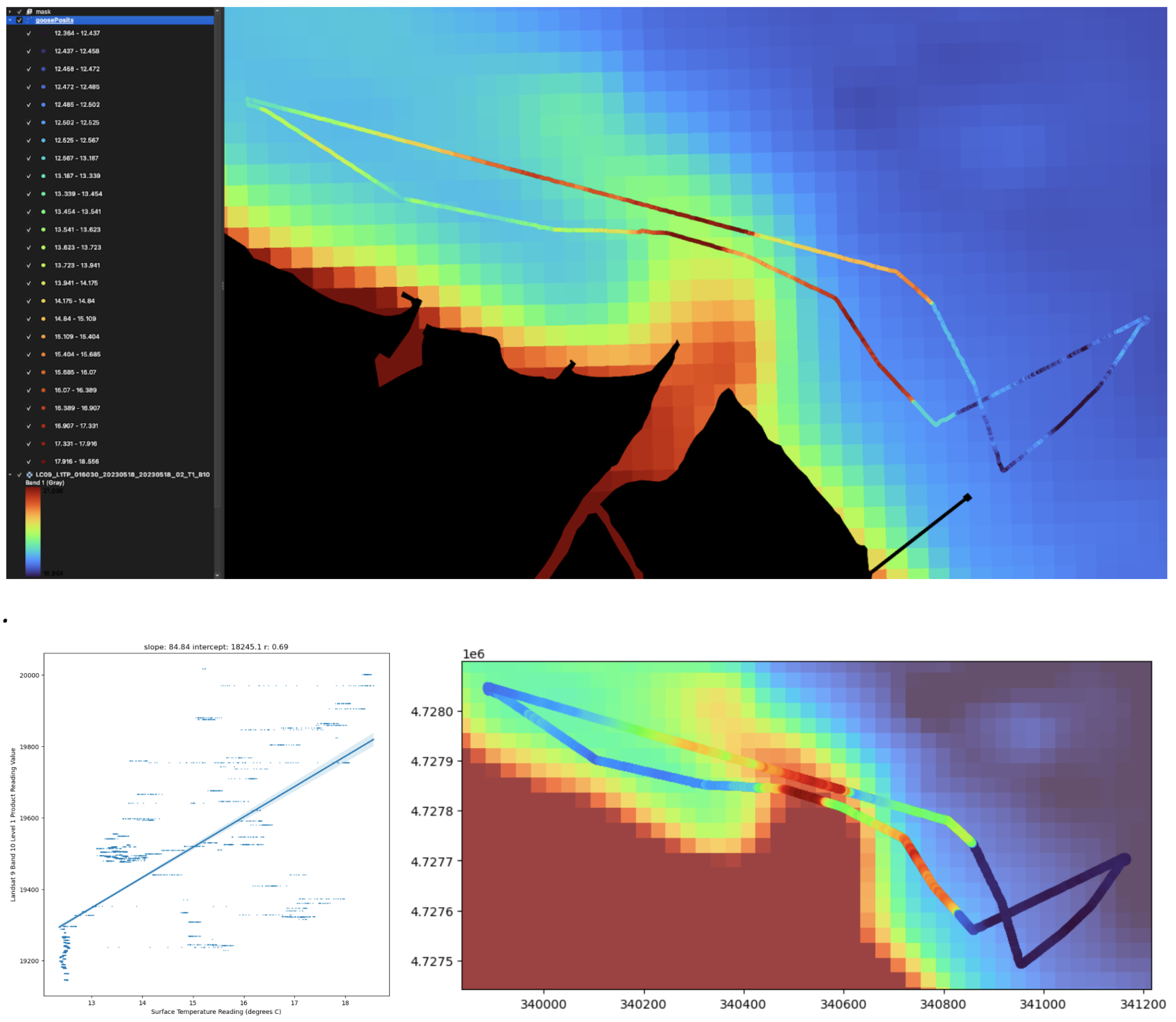}
  \caption{Satellite imagery raster of a thermal plume, taken three hours before the GANDER autonomous surface platform collected the overlaid track of temperature readings. Sensed emissivity as compared to surface measurements taken by an autonomous temperature platform as ground truth, compared spatially and as a scatter plot.}
\end{figure}

During a two hour “mission” in May, 2023, scheduled for the same day as a Landsat flyover of Lake Seneca, GANDER gathered over 4000 usable temperature points across the thermal waste plume. We deployed GANDER from a private dock over a kilometer from the plant's cooling water intake pipe, and the platform navigated autonomously, recording temperature approximately once a second (1Hz). As implied by its name, Landsat is primarily used for measuring land surface temperatures. Various algorithms exist for extrapolating water temperature, which we compared to results of the mobile platform sensor readings for purposes of calibration and corroboration. Findings demonstrated that while absolute values disagree, they clearly correlate the scale and degree of the waste plume. 

GANDER was a feasibility prototype built to demonstrate that measurements using citizen-friendly inexpensive, off-the-shelf parts might be able to capture readings across large areas of open water that previous generations of citizen-oriented water sensing platforms could not. Readings of turbidity, salinity, and other variables could also be read in the ways we did temperature. Ultimately, by proving that the Landsat readings were proxies to ground (surface) truth readings, but for free and with adequate resolution to give a sense of thermal plumes across the lake, we paused development of GANDER, but would return to it if disagreements about discharge temperatures resurface, or if readings with significantly finer temporal or spatial resolution were necessary.

A likely next step for both Sathelper and Gander would be to attempt to correlate HAB outbreaks with recorded thermal plumes. Doing so goes beyond the expertise of our team, and would require collaboration with experts in freshwater biology. That said, no one sensor, dataset, or platform will single-handedly solve a technically contested dispute like the one around Greenidge, but neither should approaches that can provide data in the interests of communities be dismissed for that reason. Time and again, members of communities faced with environmental challenges find themselves becoming not only involuntary activists, but in many cases having to become literate about technologies and techniques they never expected or wanted to know. Citizen- or community- centered technical platforms can help to scaffold this learning process, help to introduce the technical nature of problems, be used to strategize more technically sophisticated future steps, and grant entrée into technical exchanges between regulators and industry.

\section{DISCUSSION \& REFLECTIONS} 

The projects described above provide important lessons and resources for other communities and HCI researchers engaged in projects of environmental contestation, and have begun to be taken up in other contexts.  Beyond Seneca Lake, SatHelper has been used to identify and document the thermal plumes of a bitcoin mining operation in a "remediated" superfund site neighboring a Native American reservation community elsewhere in New York. This has led to more detailed investigation and evaluation of the local effects of heat pollution by grass-roots Indigenous and environmental activists and partnered university researchers. Members of the project team are also currently applying features of DEC\_Counts to two other case studies, 1) a community controversy surrounding the expansion of a landfill; and 2) a socio-ecological impact study regarding the construction of a hydropower dam.  To support further sharing, adoption, and adaptation – an important commitment of our method – code and more detailed design and use specifications for each of the interventions can be found at: [WEBSITE REMOVED FOR PURPOSES OF REVIEW.]  

An example of how these same platforms might be extensible in different spaces of contention can be found in an unexpected discovery by one of our researchers when working on a different project with a Native American community facing environmental justice challenges 200 miles away. Curious about the thermal impact of a plant near this community, they reached to SatHelper to get a sense of its discharge, noticed that next to their area of interest, the empty brownfield of a "remediated" superfund site was emitting heat. This has led to a more detailed investigation and evaluation of the cause of the heat, led by grass-roots Indigenous, environmental activists and partnered university researchers, with the current hypothesis that in-ground bacteria are consuming the ample residual polychlorinated biphenyls (PCBs) in the site; the waste byproduct of such metabolization has been shown elsewhere to result in even more volatile PCBs. Similarly, we are applying certain features of DEC\_Counts to two other case studies, 1) a community controversy surrounding the expansion of a landfill and 2) a socio-ecological impact study regarding the construction of a hydropower dam. 

This work also posed distinct challenges and limits.  One of these concerned the deep epistemic divide between corporate practices and the (unresourced) publics and (under-resourced) regulators seeking to check and monitor these activities. While the tools provided went some small distance towards bridging this gap, important disparities remain. A second challenge involved timing: notably the difficulty of generating knowledge able to feed regulatory and political debates in a timely way; and conversely, the frustration of repeated delays and postponements in hard-won impacts (as evidenced in the zombie permit discussion above). A final challenge concerned the weight of evidence required in the spheres of regulatory science and the legal system more generally. Put simply, the community was looking for data ‘hard’ enough to stop a Bitcoin’ mining operation; even after the interventions above, this standard proved elusive to achieve.  

But as this last point begins to suggest, disparities between official and activist accounts of the Greenidge situation extended beyond matters of fact such as temperature levels, public comment statistics, and regulatory deadlines. At issue was something akin to the “rules of the game” being played by each side. In spite of what activists saw as a clear interdependence of technical and political concerns in the Greenidge case, the official narrative treated these concerns as mutually exclusive so as to more credibly present regulatory decisions as entirely procedural or technical in nature. It was not lost on activists, however, that the handling of technical questions such as the practical implications of an expired permit, or procedural questions like the proper assessment of public comments, typically erred on the side of maintaining business-as-usual for Greenidge.

Between the sheer political and technical complexity of the issue and active obfuscation on the part of industrial actors, we neither expected nor were surprised by the lack of a quick technical fix, or “smoking gun” data point capable (on its own) of forcing a suspension of bitcoin mining. The results of DEC\_Counts did not halt Greenidge's operations, and the data outputs of GANDER have (to date) been professionally ignored by regulatory agencies and litigants. But any assessment of the potential impact and value of these computational interventions must extend beyond the immediate question of ‘solutions’ or answers, to consider wider and longer-term factors and influences: for example, the fostering of community voice and expertise; the building of new collaborative relations between groups and community members previously disconnected or in conflict; and the establishment of new (if still evolving) tools, processes and infrastructures for responding to the uneven politics of expertise found around situations of local environmental controversy. While harder to track (and remaining very much in progress at time of writing) we believe that it is at the level of these wider processes – and not at the level of design per se – that the most important lessons for HCI may lie.  

\subsection{Repertoires of Contention}

Beyond the level of design, the interventions above suggest a different ambition and goal for HCI work: namely, its ability to serve as a prop or tool of democratic contention: an agent of (good) conflict, rather than technical solutions. Here we take insight from what sociologists Charles Tilly and Sidney Tarrow have called “repertoires of contention”, or “the means that ordinary people use to make consequential collective claims” \cite{tilly_democracy_2007}. As Tilly and Tarrow and others in the ‘contentious politics’ tradition note, there are remarkable consistencies of form that social and political movements throughout history have adopted: the campaign, the rally, the petition, the public meeting \cite{tilly_social_2019}. These forms adapt and change with new forms of politics (many pre-date representational democracies, relying on pressuring a regent rather than a congress); occasionally new ones may be invented to take advantage of  developments in media (as in the role of newspapers in the abolition of slavery).

This borrowing and ‘genre-fication’ of contentious activity is part of what establishes the creativity but also limits of conflict and dissent over time, making matters of concern legible and tractable in the eyes of wider groups and institutions. Repertoires of contention articulate concrete issues and concerns into forms that can be communicated to and absorbed by a broader public. And in cases like the ones reviewed above, they may claim and re-work the forms and tropes of science and technology themselves, suggesting new lessons in what ‘good’ science and ‘good’ design might mean.

Repertoires of contention reverse a presumption towards integration and cooperation that has sometimes framed and delimited HCI work, suggesting instead that conflict (rather than cooperation) can be the point and goal of computational design – a point picked up in related work in democratic theory \cite{mouffe_deliberative_1999, connolly_pluralism_2005} and some variants of critical or adversarial design \cite{dantec_designing_2016, disalvo_adversarial_2015}. If repertoires of contention have implications for the analysis of social movements, they may also hold lessons about collaboration, community engagement, and the work of technologists themselves. In the Greenidge space, we saw contention between an \textit{ad hoc} group of citizens and a company using racks of Graphics Processing Units (GPUs) to engage in brute-force random number guessing for profit (with the former described by all parties involved as “activist,” and the latter as industrial and ‘technical’). Our intervention was also about questioning and rethinking this framing. Working creatively within these repertoires of contention demanded new ways of thinking for some members of our study team, some of which challenged the practices and presumptions of our own prior work (but also potentially HCI research more generally).

First, we learned to account for community-based agendas in scholar/activist collaborations. In our research framework, acknowledging our envelopment in the repertoires of contention, community activists set the agenda and elaborate on what their sources and metrics of concern are. Members of the research team were situated with, located near, accountable and in many ways beholden to our collaborators. While taking a definite 'side' in opposition to Greenidge and energy-intensive cryptocurrency mining generally, the environmental impacts and supposed benefits of which we find problematic at best.\footnote{Several of us actively participated (and continue to) be a part of the National Coalition Against Crypto-mining (detailed above).} 

Accounting for community-based agendas oftentimes took a reflexive turn and unveiled uncomfortable tensions these often separate subject positions entailed, producing some mixed sympathies and ambivalent collaborations. For example, through the projects above our relation to the DEC was hard to conceptualize in any neat and tidy way. On the one hand we sympathized with their lack of resources and were interested in seeing their nominal mission carried out (as, of course, were the many well-intentioned DEC employees). On the other hand, we no doubt further contributed to their already unmanageable workload with our public records requests, and our publication of the public comments data could be easily seen as a kind of public shaming exercise. Other challenges and ambivalences concerned the stance and standpoint of the research team itself within our roles as academic researchers. Throughout this work, we have been reflective and critical of our own objectives and desired outcomes - as it is often difficult to deconstruct academic orientations and pursuance of scientific innovation, scalability and general knowledge production. This is especially the case when one's own university is actively involved in a consortium to build a massive regional data center, one that will almost surely reproduce some of Greenidge's impacts, in order to power academic AI research (in part as an admirable alternative to the AI industry's domination of the field). That acknowledged, we centered our goals towards utilizing our university affiliation and resources to build systems that expand on an agenda fundamentally set by the needs of our community partners on the one hand, and, on the other, the hamstrung efforts and resources of government regulators (e.g., the DEC) charged with adjudicating environmental controversy.

Second, these interventions posed new challenges in navigating the visibility of HCI and design work – and in particular think across contexts where our work was meant to be public and center stage (like DEC Counts) and those where it was almost literally meant to disappear (like GANDER).  As seen in our studies, some of this will involve a certain kind of politically-inflected articulation work \cite{star_layers_1999}: for example, the work to make nominally public satellite data available to wider classes of non-expert users; or to make public comment periods demanded under the notice and input requirements of regulatory process more effective and transparent barometers of public opinion (for both regulators and for the “publics” themselves). These kinds of activities are in turn central to the work of turning publics on paper into publics in practice - one more way in which infrastructures (and \textit{specific} infrastructures) of the kind long-studied in HCI are central and not incidental to the wider social orders in which they live.

\section{CONCLUSION} 
Through our ongoing engagement with the Greenidge Bitcoin-mining power plant controversy, we have explored the ways in which HCI-inflected interventions can foster productive collaborations with local activists. Rather than presenting stakeholders with a predetermined research program oriented around specific matters of fact, we expanded upon the present work of community and civic science and data activism research. The direction and tempo of the work was largely set by activists’ struggles with a regulatory process, tightly coupled government and industrial interactions. Central to this project is the production of new computationally-supported \textit{repertoires of contention} - a concept and practice that seeks to bring HCI design efforts, community engagement, and democratic theory and practice together in new, productive and more open-ended ways.   

In the case of our DEC\_Counts intervention, for example, we noted widely held frustrations with how a deluge of public sentiment–virtually all of which demanded a quick and decisive response–had become an \textit{excuse} for further delaying the regulatory process. By obtaining, analyzing, and reporting on the content of the public comments, we were able to render popular sentiment into publicly legible through journalistic channels, converting democratic expression from yet another dataset into a ready-to-hand political tool. In alignment with thermal pollution measurement efforts already undertaken by community members, we began developing a broad toolset of low-cost, low-profile monitoring techniques suitable to the legal-political and practical limitations faced by people living near the plant. These computational techniques represent ongoing experiments in technologically mediated community organizing that include but extend beyond measurement. As researchers, we strategically work and position ourselves with organizers to create interventions that serve as computational apparatuses within repertoires of contention. With FOIL requests still outstanding, adjudicatory processes underway, and data still to collect, our involvement in the Greenidge case is still very much in progress. 

\begin{acks}
[Redacted]
\end{acks}

\bibliographystyle{ACM-Reference-Format}
\bibliography{FINAL-BIB-CHI}

@article{meng_grassroots_2018,
	title = {Grassroots resource mobilization through counter-data action},
	volume = {5},
	issn = {2053-9517},
	url = {https://doi.org/10.1177/2053951718796862},
	doi = {10.1177/2053951718796862},
	language = {en},
	number = {2},
	urldate = {2022-12-08},
	journal = {Big Data \& Society},
	author = {Meng, Amanda and DiSalvo, Carl},
	month = jul,
	year = {2018},
	pages = {2053951718796862},
}

@misc{zafiris_marina_greenidge_nodate,
	title = {Greenidge {Power} {Network}},
	url = {https://littlesis.org/oligrapher/7566-greenidge-power-network},
	urldate = {2023-06-08},
	author = {{Zafiris, Marina} and {Marshall, Owen}},
}

@techreport{asa_analysis__communication_inc_greenidge_2022,
	address = {921 Pike Street, PO Box 303 Lemont, PA 16851},
	title = {{GREENIDGE} {THERMAL} {STUDY} {FINAL} {INTERIM} {REPORT}},
	url = {https://treichlerlawoffice.com/water/greenidge/5-25-22%20Final_Interim_Report.pdf},
	author = {{ASA Analysis \& Communication, Inc.}},
	year = {2022},
}

@article{latour_why_nodate,
	title = {Why {Has} {Critique} {Run} out of {Steam}? {From} {Matters} of {Fact} to {Matters} of {Concern}},
	language = {en},
	author = {Latour, Bruno},
}

@misc{mantius_greenidge_2021,
	title = {Greenidge {Wows} {Wall} {Street} {With} {Story} of ‘{Carbon} {Neutral}’ {Bitcoin} {Mining}; {But} {Is} {It} {Only} ‘{Window} {Dressing}?’},
	shorttitle = {Greenidge {Wows} {Wall} {Street} {With} {Story} of ‘{Carbon} {Neutral}’ {Bitcoin} {Mining}; {But} {Is} {It} {Only} ‘{Window} {Dressing}?},
	url = {https://waterfrontonline.blog/2021/08/26/greenidge-wows-wall-street-with-story-of-carbon-neutral-bitcoin-mining-but-is-it-little-more-than-window-dressing/},
	language = {en},
	urldate = {2023-06-29},
	author = {Mantius, Peter},
	month = aug,
	year = {2021},
	note = {Publication Title: Water Front- Peter Mantius},
}

@misc{bill_meyer_aig_2009,
	title = {{AIG} gave {N}.{Y}. {Democrats} \$100,000 just before {Gov}. {David} {Paterson} intervened to help insurer},
	url = {https://www.cleveland.com/nation/2009/03/aig_gave_ny_democrats_100000_j.html},
	language = {en},
	urldate = {2023-06-29},
	author = {Bill Meyer, The Plain Dealer},
	month = mar,
	year = {2009},
	note = {Publication Title: cleveland},
}

@article{coplan_zombie_2005,
	title = {Of {Zombie} {Permits} and {Greenwash} {Renewal} {Strategies}: {Ten} {Years} of {New} {York}'s {So}-{Called} "{Environmental} {Benefit} {Permitting} {Strategy}"},
	shorttitle = {Of {Zombie} {Permits} and {Greenwash} {Renewal} {Strategies}},
	url = {https://digitalcommons.pace.edu/lawfaculty/357},
	journal = {Elisabeth Haub School of Law Faculty Publications},
	author = {Coplan, Karl},
	month = jan,
	year = {2005},
}

@misc{mantius_cuomo_2018,
	title = {Cuomo {Administration} {Rushed} {Greenidge}’s {Restart} — {But} {Was} {It} {Legal}?},
	url = {https://waterfrontonline.blog/2018/11/13/cuomo-administration-rushed-greenidges-restart-but-was-it-legal/},
	language = {en},
	urldate = {2023-06-30},
	author = {Mantius, Peter},
	month = nov,
	year = {2018},
	note = {Publication Title: Water Front- Peter Mantius},
}

@inproceedings{jackson_infrastructure_2013,
	address = {New York, NY, USA},
	series = {{CHI} '13},
	title = {Infrastructure and vocation: field, calling and computation in ecology},
	isbn = {978-1-4503-1899-0},
	shorttitle = {Infrastructure and vocation},
	url = {https://doi.org/10.1145/2470654.2481397},
	doi = {10.1145/2470654.2481397},
	urldate = {2023-07-14},
	booktitle = {Proceedings of the {SIGCHI} {Conference} on {Human} {Factors} in {Computing} {Systems}},
	publisher = {Association for Computing Machinery},
	author = {Jackson, Steven J. and Barbrow, Sarah},
	month = apr,
	year = {2013},
	pages = {2873--2882},
}

@article{raddick_galaxy_2010,
	title = {Galaxy {Zoo}: {Exploring} the {Motivations} of {Citizen} {Science} {Volunteers}},
	volume = {9},
	issn = {15391515},
	shorttitle = {Galaxy {Zoo}},
	url = {http://arxiv.org/abs/0909.2925},
	doi = {10.3847/AER2009036},
	number = {1},
	urldate = {2023-07-14},
	journal = {Astronomy Education Review},
	author = {Raddick, M. Jordan and Bracey, Georgia and Gay, Pamela L. and Lintott, Chris J. and Murray, Phil and Schawinski, Kevin and Szalay, Alexander S. and Vandenberg, Jan},
	month = dec,
	year = {2010},
}

@article{curtis_motivation_2015,
	title = {Motivation to {Participate} in an {Online} {Citizen} {Science} {Game}: {A} {Study} of {Foldit}},
	volume = {37},
	issn = {1075-5470},
	shorttitle = {Motivation to {Participate} in an {Online} {Citizen} {Science} {Game}},
	url = {https://doi.org/10.1177/1075547015609322},
	doi = {10.1177/1075547015609322},
	language = {en},
	number = {6},
	urldate = {2023-07-14},
	journal = {Science Communication},
	author = {Curtis, Vickie},
	month = dec,
	year = {2015},
	pages = {723--746},
}

@inproceedings{eveleigh_i_2013,
	address = {New York, NY, USA},
	series = {Gamification '13},
	title = {“{I} want to be a captain! {I} want to be a captain!”: gamification in the old weather citizen science project},
	isbn = {978-1-4503-2815-9},
	shorttitle = {“{I} want to be a captain! {I} want to be a captain!”},
	url = {https://dl.acm.org/doi/10.1145/2583008.2583019},
	doi = {10.1145/2583008.2583019},
	urldate = {2023-07-14},
	booktitle = {Proceedings of the {First} {International} {Conference} on {Gameful} {Design}, {Research}, and {Applications}},
	publisher = {Association for Computing Machinery},
	author = {Eveleigh, Alexandra and Jennett, Charlene and Lynn, Stuart and Cox, Anna L.},
	month = oct,
	year = {2013},
	pages = {79--82},
}

@article{von_ahn_games_2006,
	title = {Games with a purpose},
	volume = {39},
	issn = {1558-0814},
	doi = {10.1109/MC.2006.196},
	number = {6},
	journal = {Computer},
	author = {von Ahn, L.},
	month = jun,
	year = {2006},
	pages = {92--94},
}

@article{mckinley_citizen_2017,
	series = {The role of citizen science in biological conservation},
	title = {Citizen science can improve conservation science, natural resource management, and environmental protection},
	volume = {208},
	issn = {0006-3207},
	url = {https://www.sciencedirect.com/science/article/pii/S0006320716301963},
	doi = {10.1016/j.biocon.2016.05.015},
	language = {en},
	urldate = {2023-07-14},
	journal = {Biological Conservation},
	author = {McKinley, Duncan C. and Miller-Rushing, Abe J. and Ballard, Heidi L. and Bonney, Rick and Brown, Hutch and Cook-Patton, Susan C. and Evans, Daniel M. and French, Rebecca A. and Parrish, Julia K. and Phillips, Tina B. and Ryan, Sean F. and Shanley, Lea A. and Shirk, Jennifer L. and Stepenuck, Kristine F. and Weltzin, Jake F. and Wiggins, Andrea and Boyle, Owen D. and Briggs, Russell D. and Chapin, Stuart F. and Hewitt, David A. and Preuss, Peter W. and Soukup, Michael A.},
	month = apr,
	year = {2017},
	pages = {15--28},
}

@article{peters_use_2015,
	title = {The use and value of citizen science data in {New} {Zealand}},
	volume = {45},
	issn = {0303-6758},
	url = {https://doi.org/10.1080/03036758.2015.1051549},
	doi = {10.1080/03036758.2015.1051549},
	number = {3},
	urldate = {2023-07-14},
	journal = {Journal of the Royal Society of New Zealand},
	author = {Peters, MA and Eames, C and Hamilton, D},
	month = jul,
	year = {2015},
	pages = {151--160},
}

@article{phillips_engagement_2019,
	title = {Engagement in science through citizen science: {Moving} beyond data collection},
	volume = {103},
	copyright = {© 2019 Wiley Periodicals, Inc.},
	issn = {1098-237X},
	shorttitle = {Engagement in science through citizen science},
	url = {https://onlinelibrary.wiley.com/doi/abs/10.1002/sce.21501},
	doi = {10.1002/sce.21501},
	language = {en},
	number = {3},
	urldate = {2023-07-14},
	journal = {Science Education},
	author = {Phillips, Tina B. and Ballard, Heidi L. and Lewenstein, Bruce V. and Bonney, Rick},
	year = {2019},
	pages = {665--690},
}

@inproceedings{aoki_environmental_2017,
	address = {New York, NY, USA},
	series = {{CHI} '17},
	title = {Environmental {Protection} and {Agency}: {Motivations}, {Capacity}, and {Goals} in {Participatory} {Sensing}},
	isbn = {978-1-4503-4655-9},
	shorttitle = {Environmental {Protection} and {Agency}},
	url = {https://dl.acm.org/doi/10.1145/3025453.3025667},
	doi = {10.1145/3025453.3025667},
	urldate = {2023-07-14},
	booktitle = {Proceedings of the 2017 {CHI} {Conference} on {Human} {Factors} in {Computing} {Systems}},
	publisher = {Association for Computing Machinery},
	author = {Aoki, Paul and Woodruff, Allison and Yellapragada, Baladitya and Willett, Wesley},
	month = may,
	year = {2017},
	pages = {3138--3150},
}

@article{gabrys_programming_2014,
	title = {Programming {Environments}: {Environmentality} and {Citizen} {Sensing} in the {Smart} {City}},
	volume = {32},
	issn = {0263-7758},
	shorttitle = {Programming {Environments}},
	url = {https://doi.org/10.1068/d16812},
	doi = {10.1068/d16812},
	language = {en},
	number = {1},
	urldate = {2023-07-14},
	journal = {Environment and Planning D: Society and Space},
	author = {Gabrys, Jennifer},
	month = feb,
	year = {2014},
	pages = {30--48},
}

@book{ensemble_electronic_1996-1,
	title = {Electronic {Civil} {Disobedience} and {Other} {Unpopular} {Ideas}},
	publisher = {Autonomedia},
	author = {Ensemble, Critical Art},
	year = {1996},
}

@book{dunbar-hester_hacking_2019,
	title = {Hacking {Diversity}:{The} {Politics} of {Inclusion} in {Open} {Technology} {Cultures}},
	isbn = {978-0-691-18207-0},
	url = {https://press.princeton.edu/books/hardcover/9780691182070/hacking-diversity},
	language = {en},
	urldate = {2023-07-14},
	author = {Dunbar-Hester, Christina},
	month = dec,
	year = {2019},
}

@book{dewey_public_2012,
	title = {The {Public} and {Its} {Problems}: {An} {Essay} in {Political} {Inquiry}},
	isbn = {978-0-271-05569-5},
	shorttitle = {The {Public} and {Its} {Problems}},
	url = {https://www.jstor.org/stable/10.5325/j.ctt7v1gh},
	urldate = {2023-07-14},
	publisher = {Penn State University Press},
	author = {Dewey, John},
	editor = {Rogers, Melvin L.},
	year = {2012},
}

@book{tilly_democracy_2007,
	address = {Cambridge},
	title = {Democracy},
	isbn = {978-0-521-87771-8},
	url = {https://www.cambridge.org/core/books/democracy/26A098816E37F7FB60873348D1BBEBD3},
	urldate = {2023-07-14},
	publisher = {Cambridge University Press},
	author = {Tilly, Charles},
	year = {2007},
	doi = {10.1017/CBO9780511804922},
}

@book{tilly_social_2019,
	address = {New York},
	title = {Social {Movements}, 1768-2004},
	isbn = {978-1-315-63206-3},
	publisher = {Routledge},
	author = {Tilly, Charles},
	month = jan,
	year = {2019},
	doi = {10.4324/9781315632063},
}

@article{schrock_civic_2016,
	title = {Civic hacking as data activism and advocacy: {A} history from publicity to open government data},
	volume = {18},
	issn = {1461-4448},
	shorttitle = {Civic hacking as data activism and advocacy},
	url = {https://journals.scholarsportal.info/details/14614448/v18i0004/581_chadaafptogd.xml},
	doi = {10.1177/1461444816629469},
	number = {4},
	urldate = {2023-07-14},
	journal = {New Media \& Society},
	author = {Schrock, Andrew R},
	year = {2016},
	pages = {581--599},
}

@article{currie_conundrum_2016,
	title = {The conundrum of police officer-involved homicides: {Counter}-data in {Los} {Angeles} {County}},
	volume = {3},
	issn = {2053-9517},
	shorttitle = {The conundrum of police officer-involved homicides},
	url = {https://doi.org/10.1177/2053951716663566},
	doi = {10.1177/2053951716663566},
	language = {en},
	number = {2},
	urldate = {2023-07-14},
	journal = {Big Data \& Society},
	author = {Currie, Morgan and Paris, Britt S and Pasquetto, Irene and Pierre, Jennifer},
	month = dec,
	year = {2016},
	pages = {2053951716663566},
}

@misc{noauthor_state_nodate,
	title = {State {Pollutant} {Discharge} {Elimination} {System} ({SPDES}) {Permit} {Program} - {NYS} {Dept}. of {Environmental} {Conservation}},
	url = {https://www.dec.ny.gov/permits/6054.html},
	urldate = {2023-07-16},
}

@article{mantius_greenidges_2022,
	title = {Greenidge's warm water discharges heated {Seneca} {Lake} beyond state limits, but {DEC} waiver excuses violations},
	url = {https://www.fingerlakes1.com/2022/10/07/greenidges-warm-water-discharges-heated-seneca-lake-beyond-state-limits-but-dec-waiver-excuses-violations/},
	language = {en-US},
	urldate = {2023-07-16},
	journal = {Fingerlakes1.com},
	author = {Mantius, Peter},
	month = oct,
	year = {2022},
}

@article{mouffe_deliberative_1999,
	title = {Deliberative {Democracy} or {Agonistic} {Pluralism}?},
	volume = {66},
	issn = {0037-783X},
	url = {https://www.jstor.org/stable/40971349},
	number = {3},
	urldate = {2023-07-16},
	journal = {Social Research},
	author = {Mouffe, Chantal},
	year = {1999},
	pages = {745--758},
}

@article{millerand_metadata_nodate,
	title = {Metadata {Standards}. {Trajectories} and {Enactment} in the {Life} of an {Ontology}},
	url = {https://florencemillerand.uqam.ca/en/diffusion/publications/metadata-standards-trajectories-and-enactment-in-the-life-of-an-ontology/},
	language = {en-US},
	urldate = {2023-07-16},
	author = {Millerand, Florence},
}

@article{karasti_infrastructure_2010,
	title = {Infrastructure {Time}: {Long}-term {Matters} in {Collaborative} {Development}},
	volume = {19},
	issn = {1573-7551},
	shorttitle = {Infrastructure {Time}},
	url = {https://doi.org/10.1007/s10606-010-9113-z},
	doi = {10.1007/s10606-010-9113-z},
	language = {en},
	number = {3},
	urldate = {2023-07-16},
	journal = {Computer Supported Cooperative Work (CSCW)},
	author = {Karasti, Helena and Baker, Karen S. and Millerand, Florence},
	month = aug,
	year = {2010},
	pages = {377--415},
}

@article{steinhardt_instrumented_2018,
	title = {The {Instrumented} {Ocean}: {How} {Sensors}, {Satellites} and {Seafloor}-{Walking} {Robots} {Changed} {What} {It} {Means} to {Study} the {Sea}},
	copyright = {Attribution 4.0 International},
	shorttitle = {The {Instrumented} {Ocean}},
	url = {https://ecommons.cornell.edu/handle/1813/59744},
	doi = {10.7298/X4XP735P},
	language = {en\_US},
	urldate = {2023-07-16},
	author = {Steinhardt, Stephanie Beth},
	month = aug,
	year = {2018},
}

@misc{alvarez_evolution_2020,
	title = {Evolution of academic publications in citizen science},
	url = {https://cstrack.eu/format/reports/evolution-academic-publications-citizen-science/},
	language = {en-GB},
	urldate = {2023-07-16},
	author = {Álvarez, David Roldán},
	month = dec,
	year = {2020},
	note = {Publication Title: https://cstrack.eu/
Type: Text},
}

@article{levine_inclusion_2007,
	title = {Inclusion and exclusion: {Implications} for group processes.},
	shorttitle = {Inclusion and exclusion},
	author = {Levine, John and Kerr, Norbert},
	month = jan,
	year = {2007},
}

@article{steinke_accuracy_2017,
	title = {The accuracy of farmer-generated data in an agricultural citizen science methodology},
	volume = {37},
	issn = {1773-0155},
	url = {https://doi.org/10.1007/s13593-017-0441-y},
	doi = {10.1007/s13593-017-0441-y},
	language = {en},
	number = {4},
	urldate = {2024-01-14},
	journal = {Agronomy for Sustainable Development},
	author = {Steinke, Jonathan and van Etten, Jacob and Zelan, Pablo Mejía},
	month = jul,
	year = {2017},
	pages = {32},
}

@inproceedings{agre_toward_2006,
	title = {Toward a {Critical} {Technical} {Practice}: {Lessons} {Learned} in {Trying} to {Reform} {AI}},
	shorttitle = {Toward a {Critical} {Technical} {Practice}},
	url = {https://www.semanticscholar.org/paper/Toward-a-Critical-Technical-Practice%3A-Lessons-in-to-Agre/4fa35e09af61214eeef2fd95d8db397baafa6f90},
	urldate = {2024-01-15},
	author = {Agre, P.},
	year = {2006},
}

@inproceedings{lee_human_2006,
	address = {New York, NY, USA},
	series = {{CSCW} '06},
	title = {The human infrastructure of cyberinfrastructure},
	isbn = {978-1-59593-249-5},
	url = {https://doi.org/10.1145/1180875.1180950},
	doi = {10.1145/1180875.1180950},
	urldate = {2024-01-14},
	booktitle = {Proceedings of the 2006 20th anniversary conference on {Computer} supported cooperative work},
	publisher = {Association for Computing Machinery},
	author = {Lee, Charlotte P. and Dourish, Paul and Mark, Gloria},
	month = nov,
	year = {2006},
	pages = {483--492},
}

@article{khelifa_wildlife_2016,
	title = {Wildlife monitoring: {Lure} gamers into citizen science},
	volume = {537},
	issn = {1476-4687},
	shorttitle = {Wildlife monitoring},
	url = {https://doi.org/10.1038/537488e},
	doi = {10.1038/537488e},
	language = {en},
	number = {7621},
	urldate = {2024-01-15},
	journal = {Nature},
	author = {Khelifa, Rassim},
	month = sep,
	year = {2016},
	pages = {488--488},
}

@article{ellwood_citizen_2023,
	title = {Citizen science needs a name change},
	volume = {38},
	issn = {0169-5347},
	url = {https://www.cell.com/trends/ecology-evolution/abstract/S0169-5347(23)00057-5},
	doi = {10.1016/j.tree.2023.03.003},
	language = {English},
	number = {6},
	urldate = {2024-01-15},
	journal = {Trends in Ecology \& Evolution},
	author = {Ellwood, Elizabeth R. and Pauly, Gregory B. and Ahn, June and Golembiewski, Kate and Higgins, Lila M. and Ordeñana, Miguel A. and Konrat, Matt von},
	month = jun,
	year = {2023},
	pmid = {37088666},
	pages = {485--489},
}

@article{star_layers_1999,
	title = {Layers of {Silence}, {Arenas} of {Voice}: {The} {Ecology} of {Visible} and {Invisible} {Work}},
	volume = {8},
	issn = {1573-7551},
	shorttitle = {Layers of {Silence}, {Arenas} of {Voice}},
	url = {https://doi.org/10.1023/A:1008651105359},
	doi = {10.1023/A:1008651105359},
	language = {en},
	number = {1},
	urldate = {2024-01-16},
	journal = {Computer Supported Cooperative Work (CSCW)},
	author = {Star, Susan Leigh and Strauss, Anselm},
	month = mar,
	year = {1999},
	pages = {9--30},
}

@book{connolly_pluralism_2005,
	address = {Durham, NC},
	title = {Pluralism},
	isbn = {978-0-8223-3567-2},
	publisher = {Duke University Press},
	author = {Connolly, William E.},
	year = {2005},
}

@book{disalvo_adversarial_2015,
	title = {Adversarial {Design}},
	isbn = {978-0-262-52822-1},
	publisher = {The MIT Press},
	author = {DiSalvo, Carl},
	month = aug,
	year = {2015},
}

@book{dantec_designing_2016,
	address = {Cambridge, Massachusetts},
	title = {Designing {Publics}},
	isbn = {978-0-262-03516-3},
	language = {English},
	publisher = {The MIT Press},
	author = {Dantec, Christopher A. Le},
	month = sep,
	year = {2016},
}

@article{mantius_dec_2022,
	title = {{DEC} gives {Greenidge} time window to expand {Bitcoin} mining as air permit review continues},
	url = {http://ithacavoice.org/2022/02/dec-gives-greenidge-time-window-to-expand-bitcoin-mining-as-air-permit-review-continues/},
	language = {en-US},
	urldate = {2024-01-16},
	journal = {The Ithaca Voice},
	author = {Mantius, Peter},
	month = feb,
	year = {2022},
}

@misc{mantius_hochul_2022,
	title = {Hochul {Must} {Put} {Finger} {Lakes} {Ahead} of {Campaign} {Fundraising} and {Deny} {Air} {Permit} for {Greenidge}’s {Bitcoin} {Operation}},
	url = {https://waterfrontonline.blog/2022/06/26/hochul-must-put-finger-lakes-ahead-of-campaign-fundraising-and-deny-air-permit-for-greenidges-bitcoin-operation/},
	language = {en},
	urldate = {2024-01-16},
	author = {Mantius, Peter},
	month = jun,
	year = {2022},
	note = {Publication Title: Water Front- Peter Mantius},
}

@inproceedings{tran_situating_2024,
	address = {New York, NY, USA},
	series = {{CHI} '24},
	title = {Situating {Datasets}: {Making} {Public} {Eviction} {Data} {Actionable} for {Housing} {Justice}},
	isbn = {9798400703300},
	shorttitle = {Situating {Datasets}},
	url = {https://dl.acm.org/doi/10.1145/3613904.3642452},
	doi = {10.1145/3613904.3642452},
	urldate = {2024-09-12},
	booktitle = {Proceedings of the {CHI} {Conference} on {Human} {Factors} in {Computing} {Systems}},
	publisher = {Association for Computing Machinery},
	author = {Tran, Anh-Ton and Guo, Grace and Taylor, Jordan and Chan, Katsuki Andrew and Raymond, Elora Lee and Disalvo, Carl},
	month = may,
	year = {2024},
	pages = {1--16},
}

@article{milan_alternative_2016,
	title = {The {Alternative} {Epistemologies} of {Data} {Activism}},
	volume = {2},
	url = {https://dare.uva.nl/search?identifier=d4d05216-b10a-4f1a-985c-37eaad0b4f3f},
	doi = {10.14361/dcs-2016-0205},
	language = {en},
	urldate = {2024-09-13},
	journal = {Digital Culture \& Society},
	author = {Milan, S. and van der Velden, L.},
	year = {2016},
}

@article{hogan_data_2015,
	title = {Data flows and water woes: {The} {Utah} {Data} {Center}},
	volume = {2},
	issn = {2053-9517},
	shorttitle = {Data flows and water woes},
	url = {https://doi.org/10.1177/2053951715592429},
	doi = {10.1177/2053951715592429},
	language = {en},
	number = {2},
	urldate = {2024-09-13},
	journal = {Big Data \& Society},
	author = {Hogan, Mél},
	month = dec,
	year = {2015},
	note = {Publisher: SAGE Publications Ltd},
	pages = {2053951715592429},
}

@article{brodie_cloud_2021,
	title = {Cloud ruins: {Ericsson}'s {Vaudreuil}-{Dorion} data centre and infrastructural abandonment},
	volume = {24},
	issn = {1369-118X},
	shorttitle = {Cloud ruins},
	url = {https://doi.org/10.1080/1369118X.2021.1909099},
	doi = {10.1080/1369118X.2021.1909099},
	number = {6},
	urldate = {2024-09-13},
	journal = {Information, Communication \& Society},
	author = {Brodie, Patrick and Velkova, Julia},
	month = apr,
	year = {2021},
	note = {Publisher: Routledge
\_eprint: https://doi.org/10.1080/1369118X.2021.1909099},
	pages = {869--885},
}

@article{lally_computational_2022,
	title = {Computational parasites and hydropower: {A} political ecology of {Bitcoin} mining on the {Columbia} {River}},
	volume = {5},
	issn = {2514-8486},
	shorttitle = {Computational parasites and hydropower},
	url = {https://doi.org/10.1177/2514848619867608},
	doi = {10.1177/2514848619867608},
	language = {en},
	number = {1},
	urldate = {2024-09-13},
	journal = {Environment and Planning E: Nature and Space},
	author = {Lally, Nick and Kay, Kelly and Thatcher, Jim},
	month = mar,
	year = {2022},
	note = {Publisher: SAGE Publications Ltd STM},
	pages = {18--38},
}

\end{document}